# A density-driven first-order phase transition in liquid sulfur

Laura Henry[1], Mohamed Mezouar[1], Gaston Garbarino[1], David Sifré[1], Gunnar Weck[2] & Frederic Datchi[3]

[1]*European Synchrotron Radiation Facility (ESRF), 71, Avenue des Martyrs, Grenoble, France*

[2]*CEA, DAM, DIF, F-91297 Arpajon, France*

[3]*Institut de Minéralogie, de Physique des Milieux Condensés et de Cosmochimie (IMPMC), Sorbonne Universités - UPMC Univ. Paris 6, CNRS UMR 7590, IRD UMR 206, MNHN, 4 place Jussieu, F-75005 Paris, France*

First-order phase transitions are characterized by a discontinuous first derivative of the Gibbs free energy, so that volumes and entropies are discontinuous. Such transitions are common in the crystalline state, but extremely rare in liquid substances and experimentally evidenced in only one pure element, phosphorus. Here we report combined *in-situ* Raman scattering, X-ray diffraction and density measurements that support the existence of a first-order liquid-liquid transition in sulfur at high pressures and temperatures. The transformation involves a sharp density jump between two structurally and dynamically distinct liquids. The first-order phase transition proceeds through an initial stage of temperature induced polymerization and a final stage where the low-density liquid abruptly converts to a denser polymeric state. This unique feature is explained by competing effects of pressure and temperature.

When subjected to high pressure and high temperature conditions, most crystalline solids undergo first-order phase transitions that may result in drastic modifications of their physico-chemical properties. For example, semi-conducting silicon becomes a metal[1] and black carbon graphite transforms into transparent diamond[2]. First-order liquid-liquid phase transitions (LLPT) which separate two liquid phases of a single-component system with distinct thermodynamic, structural and dynamic properties are much less common and their existence and nature are actively debated topics since the 1990's. Such LLPTs have been predicted from computer simulations of important substances including water[3], carbon dioxide[4], carbon[5], hydrogen[6] and nitrogen[7]. Experimental evidence for a LLPT has been found for $Y_2O_3$–$Al_2O_3$[8] mixtures, water[9] and triphenyl phosphite[10] in the supercooled regime. However, the direct *in-situ* observation of a density-induced LLPT between a low-density molecular fluid and a high-density polymeric liquid[11,12] has been unambiguously evidenced in only one pure element: phosphorus. This transition is accompanied with large structural modifications and an abrupt density jump of more than 30%[13] which is a clear signature of a density-driven first-order phase transition.

Sulfur and phosphorus are neighbours in the periodic table of the elements and their pressure-temperature phase diagram exhibits important similarities. At ambient pressure, they both can be obtained in several allotropic forms based on molecular units or polymeric chains[14], most of them being metastable. The stable form of sulfur at room P and T is orthorhombic α-sulfur consisting of S8 ring-shaped molecules, while under high pressure (P)-temperature (T) conditions, the stable polymorph is a polymeric solid composed of helical chains[15,22]. At 371 K and room P, α-sulfur converts into monoclinic β-sulfur in which the S8 molecules are preserved[17]. This local molecular arrangement is conserved upon melting at 388 K and up to 432 K where the transition termed as "lambda transition" (λ-T) occurs[18,19]. The λ-T is an abrupt and reversible partial polymerization transition where a fraction of the S8 cyclic molecules open up to form long polymeric chains. Across the λ-T, the viscosity of liquid sulfur increases by two orders of magnitude within a few degrees. This transformation is also associated to subtle enthalpy and density anomalies[20] and is incomplete[21], the polymer content reaching a maximum of ~60% at the boiling point (T=718 K). This contrasts with the LLPT transition in phosphorus which proceeds through a sharp and complete pressure-induced polymerization[11,13]. At high pressures and temperatures, several P,T domains with different thermal and electrical properties have been identified in liquid sulfur[23]. A recent *in-situ* X-ray diffraction (XRD) structural study[24] evidenced the existence above 6 GPa of a purely polymeric liquid composed of long chains below 1000 K, which evolve at higher temperatures towards another polymeric form with shorter chains. *Ab-initio* molecular dynamics simulations[25] reproduced this temperature-induced chain breakage in the high pressure liquid but found no discontinuous change of density associated to this process. It is worth noting that, so far, no *in-situ* structural or vibrational studies have been conducted in the low pressure region below 1 GPa where the molecular and polymeric liquids coexist.

We performed *in-situ* Raman scattering, X-ray diffraction and absorption measurements at the beamline ID27 of the European Synchrotron Radiation Facility (ESRF)[26] to probe the structure, dynamics and density evolution of liquid sulfur in an extended pressure-temperature (P-T) region near the lambda transition. The P-T paths in the experimental phase diagram of sulfur are presented in Figure 1. In a first series of experiments, we collected combined Raman and XRD data at room pressure and high temperature across the melting point and lambda transition by heating the sample contained in a capillary (Supplementary Information S1). Then, we compared these measurements to another series of high pressure structural and vibrational data collected across the stability fields of the molecular and polymeric liquids using diamond anvil cells (Supplementary Information S2). Finally, we performed combined density and XRD measurements in the Paris-Edinburgh press (Supplementary Information S3) to unambiguously establish the nature of the observed structural transformation.

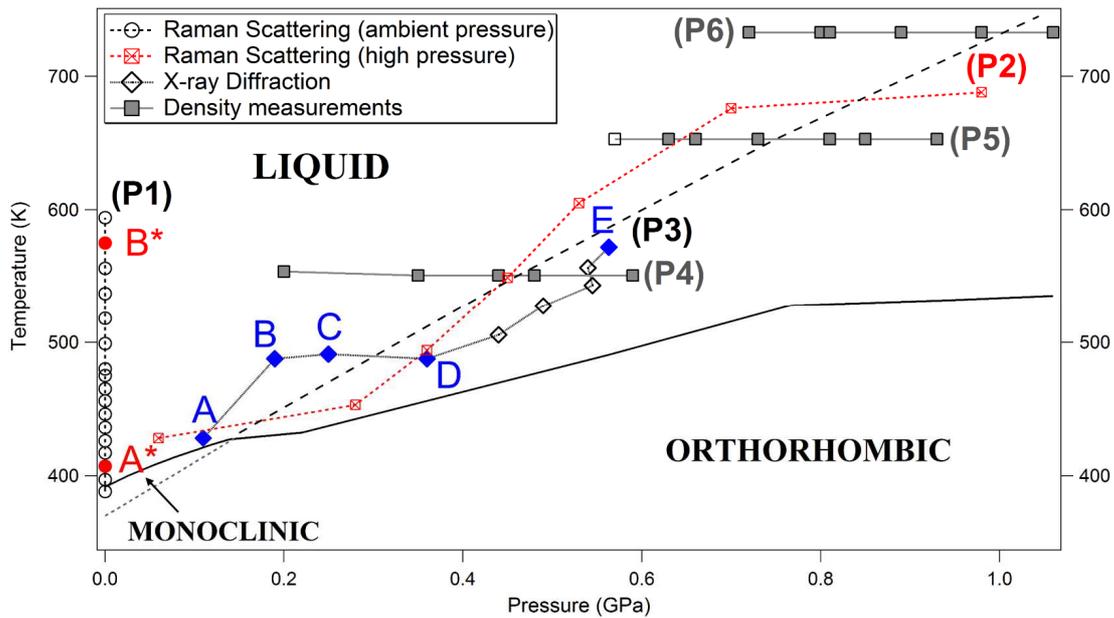

Figure 1: Experimental P,T pathways presented in a previously reported phase diagram of sulfur [15]. P1 (open circles): combined high temperature Raman and XRD experiments at ambient pressure. Selected Raman spectra and diffraction patterns are shown in Supplementary Information S1. $A^*$ and $B^*$ (red circles) correspond to the temperature conditions at which the ambient-pressure neutron data reproduced from reference 19 and presented in the insert of Figure 3b were collected. P2 (red squares): P,T pathway of the Raman experiment presented in Figure 2a. We used the Raman spectra to calculate the polymer contents C{poly} under ambient and high pressure conditions (see Figure 2b and Supplementary Information S1). P3 (open diamonds): P, T pathway of the XRD experiments presented in Figures 3a and 3b (Supplementary Information S1). The letters A, B, C, D, E (blue diamonds) indicate the P, T conditions of the selected data in Figures 3a and 3b. P4, P5, P6: isothermal pathways of the density measurements (Supplementary Information S1). The density jump presented in Figure 4 is obtained along P5. The open square in P5 indicates the P,T conditions where the density measurement after releasing the pressure was performed. The Grey dashed line is the separation line between the low and high density liquids.

As demonstrated in previous high temperature Raman scattering studies of sulfur[18], the polymer content C{poly} above the λ-T can be directly obtained from the intensity ratio of purely molecular and polymeric vibrational modes. The temperature evolution of C{poly} obtained from the Raman spectra collected along the path P1 in Figure 1 is shown in Figure 2b. In agreement with previous works[18], at ambient pressure intense molecular modes are persistent in a wide temperature domain (Supplementary Information S1) demonstrating that polymerization is incomplete (C{poly} saturates at ~60%). At high pressures, the process of transformation is radically different. Indeed, we observed a complete disappearance of the molecular modes concomitant with an intensity increase of the polymeric modes in a narrow P-T domain (pathway P2 in Figure 1). This is illustrated in Figure 2a and 2b where a typical series of Raman spectra collected at temperatures up to 670 K and pressures between 0.1 and 1 GPa and the corresponding evolution of C{poly} are presented. C{poly} reaches 100% within a small pressure interval attesting that liquid sulfur abruptly converts to a purely polymeric state. Such behavior is a clear indication for a density-driven first-order phase transition and shows that at ambient pressure, the completion of the polymerization process (λ-T) could be inhibited due to competing effects of pressure and temperature.

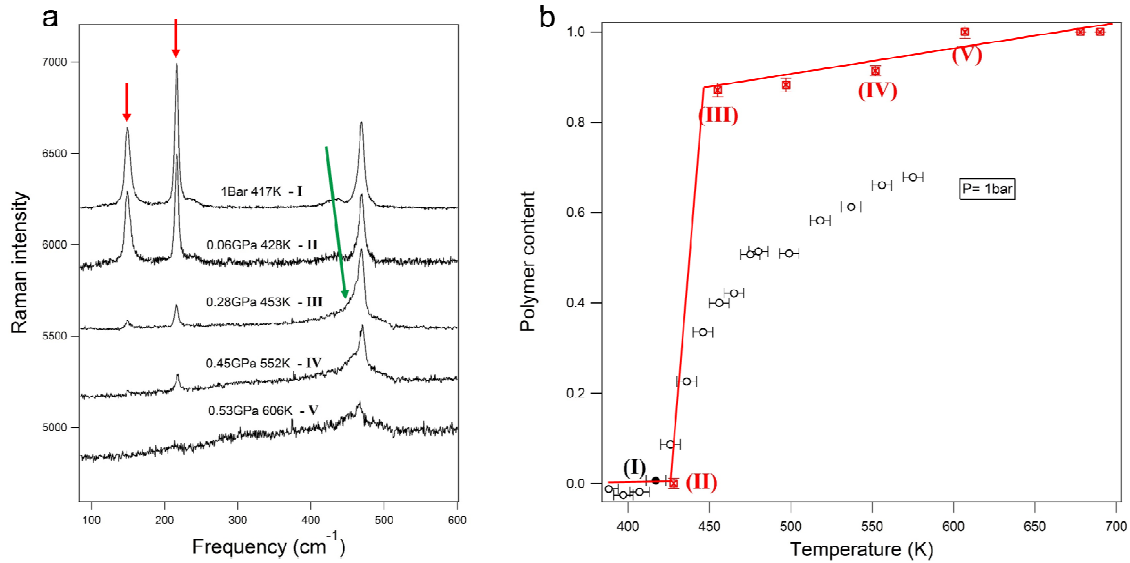

Figure 2. (a). Selected Raman spectra of liquid sulfur collected along P2 in Figure 1. The red arrows indicate the bending mode at 152 cm$^{-1}$ and breathing mode at 220 cm$^{-1}$ associated with the free S8 molecule. The Green arrow designates the shoulder on the stretching mode at 475 cm$^{-1}$, signature of the polymerization at temperatures above the lambda transition (Supplementary Information S1); (b). Evolution of the Polymer content at ambient pressure (open circle, P1 in Figure 1) and under pressure (red squares, P2 in Figure 1). The Roman numbers (I to V) indicates the correspondence between the Raman spectra in Figure 2a and polymer content in Figure 2b.

Additional critical information is provided by *in-situ* high energy resolution XRD. In contrast to phosphorus[10], the structure factor S(Q) of liquid sulfur does not exhibit a first sharp diffraction peak (FSDP). The FSDP is a manifestation of intermediate-range order on length scales larger than those typical of nearest-neighbour distances[27]. In phosphorus, it is related to medium range order correlations of the tetrahedral P4 units present in the liquid state at low pressure. The abrupt disappearance of the FSDP upon pressurization in phosphorus was used as main criterion to establish the first-order nature of the liquid-liquid transition[10]. The absence of a FSDP in liquid sulfur denotes the absence of such intermediate-range correlations between the S8 molecules. However, significant and sharp changes occur in the structure factor S(Q) and corresponding pair correlation function G(r) during the complete polymerization process. At ambient pressure, neutron diffraction experiments[19] showed that across the λ-T the S(Q) of sulfur exhibits minor changes at momentum transfer values Q higher than 3 Å$^{-1}$ attesting that the very short-range order is not affected when the S8 molecules open to form long polymeric chains. The main modifications are an intensity decrease of the first peak at Q ~ 3 Å$^{-1}$, and the broadening of the pre-peak at Q ~ 1.3 Å$^{-1}$ which almost collapses into the main peak. In real space, the most significant changes on G(r) occur at distances between 4 and 5 Å which corresponds to the third- and fourth-neighbour distribution. Similar changes occur in the S(Q) and G(r) determined from our high-pressure measurements along pathway P3 of Figure 1, as presented in Figures 3-a and 3-b. We indeed also observe, but in a more pronounced way, a drastic reduction of the intensity of the first peak at Q ~ 3 Å$^{-1}$ and a collapse of the pre-peak, while there is little modification of the S(Q) at higher Q values. More importantly and in contrast with the ambient-pressure neutron data[19] we observe a discontinuous pressure variation of $(Q_M/Q_0)^3$ where $Q_M$ is the position of the first peak maximum in the S(Q) (insert of Figure 3-a). This quantity is linearly related to density for simple liquids[29]. We also evidenced a drastic modification of the fourth-neighbour distribution (insert of Figure 3-b) as the peak of G(r) at 4.5 Å, which is persistent in the ambient-pressure neutron data[19], totally vanishes at high pressure therefore confirming a complete and sharp transformation. In agreement with our high-pressure Raman scattering data this behavior is suggestive of a first-order phase transition between two structurally distinct liquids.

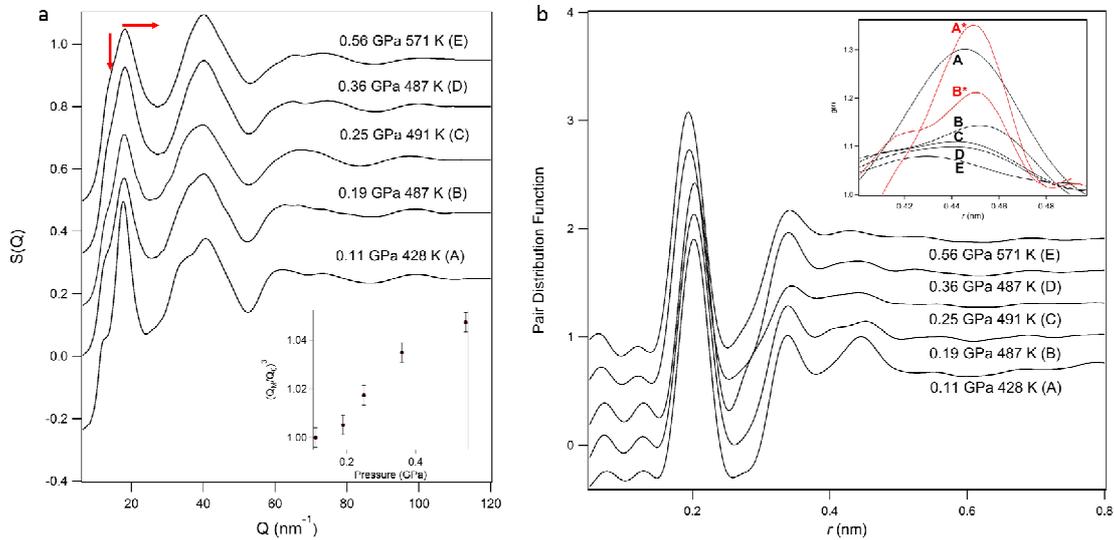

Figure 3ab. Structure factor, S(Q) and corresponding pair correlation function G(r) for liquid S at selected pressure and temperature conditions along P3 in Figure 1(Supplementary Information S2). The letters in parentheses indicate data points in Figure 1 where the measurements were performed. a. The red arrows in the left panel designate the positive shift of the first peak position and the collapse of the pre-peak upon pressurization. The discontinuous variation of the first peak position in the S(Q) is emphasized in the inset panel where the evolution along P3 of $(Q_M/Q_o)^3$ ($Q_M$ and $Q_o$ are the first peak position in the S(Q) under pressure and at 1 bar, respectively) is shown. b. Large changes on G(r) occur only at distances between 4 and 5 Å. An enlargement of the third peak of G(r) together with ambient-pressure neutron data[19] (A$^*$ and B$^*$ in Figure 1) are shown in the inset panel. In contrast to the neutron data, we observed a complete disappearance of the peak at 4.5 Å along P3.

To unambiguously confirm the first-order nature of the transformation, we performed direct density measurements in the Paris-Edinburgh press using the X-ray absorption method developed by Y. Katayama and collaborators[28] along three isothermal pathways (P4, P5, P6 in Figure 1). The precision of the density measured by this method is better than 1% (Supplementary Information S3). XRD data were also collected at every pressure point to confirm that the density measurements were conducted on a fully molten sample. As shown in Figure 4, at 653 K we observe a large density jump of 7% at 0.6 GPa over a very narrow pressure domain of 0.07 GPa, confirming the first-order nature of the polymerization transition. In addition, a negative density jump has been observed by releasing the pressure along the three isothermal paths attesting that the transformation is fully reversible.

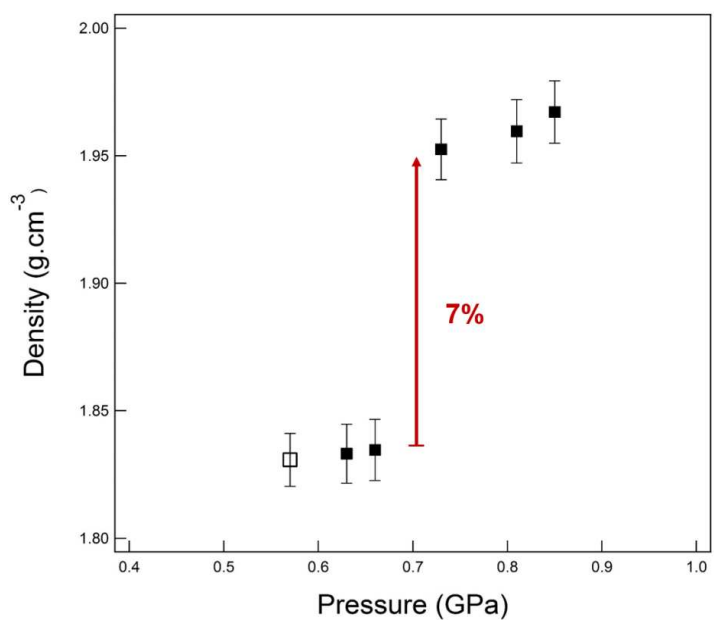

Figure 4. Pressure evolution of the liquid density at T=653 K collected along the isothermal pathway P5 in Figure 1. Black and open squares correspond to density measurements upon compression and after pressure release, respectively. We observed a density jump of 7% at 0.66 GPa in a very small pressure interval of 0.07 GPa from a low to a high density form of liquid sulfur.

The main difference with the LLPT observed in phosphorus resides in the existence of a finite P,T region near the λ-T where pressure and temperature effects are competing. At ambient pressure, the S8 molecules might open and connect with first neighbors when sufficient thermal energy is given to the system to reach, thanks to thermal motion, a critical intermolecular separation that promotes charge transfer between the S8 units. As temperature further increases, thermal expansion counteracts against this effect by separating the molecules and preventing full polymerization. Under pressure, the thermal expansion is reduced, enabling to reach the critical density required for the complete polymerization. Therefore, the polymerization in liquid sulfur proceeds through an initial stage of temperature induced partial polymerization and a final stage where the low-pressure partially polymeric liquid abruptly converts to a denser purely polymeric state. It may appear surprising that no jump in density occurs across the λ-T at ambient P while a LLPT with a discontinuous jump is observed under pressure. We think this could be due to the fact that the critical density is reached dynamically at ambient P, through thermal motion, while under pressure it is reached statically, through a bulk reduction of volume. It is likely that the kinetics of the transition is faster in the latter case giving rise to a sharp discontinuous change of density. This behavior is unique among all the elements of the periodic table and is compatible with a ''two-state'' model[30] where an intermediate P,T domain exists between the low and high-density species due to the competition between pressure and temperature effects. This very peculiar behavior of liquid sulfur should stimulate further theoretical and experimental studies.

Acknowledgments

The authors acknowledge the European Synchrotron Radiation Facility for provision of synchrotron radiation on beamline ID27 and the Agence Nationale de la Recherche for financial support under Grant No. ANR 13-BS04-0015 (MOFLEX).

References

1. Minomura S., & Drickamer, H.G., Pressure induced phase transitions in silicon, germanium and some III–V compounds. *J. Phys. Chem. Solids* 23, 451 (1962)

2. Bundy, F. P., Hall, H. T., Strong, H. M. & Wentorf, R. H. Man made diamonds. *Nature* **176,** 51–55 (4471).

3. Harrington, S., Zhang, R., Poole, P. H., Sciortino, F. & Stanley, H. E. Liquid-liquid phase transition: Evidence from simulations. *Phys. Rev. Lett.* **78**, 2409–2412 (1997).

4. Boates, B., Teweldeberhan, A. M., & Bonev, S. A. Stability of dense liquid carbon dioxide. *Proc. Natl. Acad. Sci. U. S. A.* **109**, 14808–14812 (2012).

5. Glosli, J. N. & Ree, F. H. Liquid-liquid phase transformation in carbon. *Phys. Rev. Lett.* **82**, 4659– 4662 (1999).

6. Morales, M. A., Pierleoni, C., Schwegler, E. & Ceperley, D. M. Evidence for a first-order liquid-liquid transition in high-pressure hydrogen from ab initio simulations. *Proc. Natl. Acad. Sci. USA* **107**, 12799 (2010).

7. Boates, B. & Bonev, S. First-order liquid-liquid phase transition in compressed nitrogen. *Phys. Rev. Lett.* **102**, 015701 (2009)

8. Aasland, S. & McMillan, P. F. Density-driven liquid-liquid phase separation in the system Al2O3-Y2O3. *Nature* **369**, 633-636 (1994).

9. Mishima, O. & Stanley, H. E. The relationship between liquid, supercooled and glassy water. *Nature* **396**, 329-335 (1998).

10. Tanaka H., Hurita R. & Mataki H. PRL 92, Liquid-liquid transition in the molecular liquid triphenyl phosphite. Phys. Rev. Lett. **92,** 025701-4 (2004)

11. Katayama, Y., Mizutani, T., Utsumi, W., Shimomura, O., Yamakata, M., & Funakoshi, K. A first-order liquid-liquid phase transition in phosphorus. *Nature*, **403**, 170–173, (2000).

12. Monaco, G., Falconi, S., Crichton, W. A., & Mezouar, M. Nature of the first-order phase transition in fluid phosphorus at high temperature and pressure. *Phys. Rev. Lett.* **90**, 255701 (2003)

13. Katayama, Y., Inamura, Y., Mizutani, T., Yamakata, M., Utsumi, W. & Shimomura, O. *Science* **306**, 848-851 (2004)

14. Steudel, R. & Eckert, B. Solid sulfur allotropes. *Top. Curr. Chem.* **230**, 1–79 (2003).

15. Crapanzano, L., Crichton, W. A., Monaco, G., Bellissent, R. & M. Mezouar. Alternating sequence of ring and chain structures in sulphur at high pressure and temperature. *Nature Materials* 4, 550–552 (2005).

16. Kikegawa, T. *et al.* Synchrotron-radiation study of phase transitions in phosphorus at high pressures and temperatures. *J. Appl. Crystallogr.* **20**, 406–410 (1987).

17. Templeton, L. K., Templeton, D. H. & Zalkin, A. Crystal-structure of monoclinic sulfur. *Inorganic Chemistry* **15**, 1999–2001 (1976).

18. Kalampounias, A. G., Kastrissios, D. T. & Yannopoulos, S. N. Structure and vibrational modes of sulfur around the lambda-transition and the glass transition. *Journal of Non Crystalline Solids* **326**, 115–119 (2003).

19. Bellissent, R., Descotes, L., Boué, F. & Pfeuty, P. Liquid sulfur: Local-order evidence of a polymerization transition. *Physical Review B*, **41**, 2135–2138, (1990).

20. Sauer, G. E. & Borst, L. B. Lambda transition in liquid sulfur. *Science* **158**, 1567 (1967)

21. Steudel, R. Liquid sulfur. *Top. Curr. Chem.* **230**, 81–134 (2003).

22. Crichton, W. A., Vaughan, G. B. M. & Mezouar, M. In situ structure solution of helical sulfur at 3 GPa and 400C. *Z. Kristallogr.* **216**, 417–419 (2001).

23. Brazhkin, V. V., Popova, S. V. & Voloshin, R. N. Pressure-temperature phase diagram of molten elements: selenium, sulfur and iodine. *Physica B* **265**, 64–71 (1999).

24. Liu, L. Kono, Y., Kenney-Benson, C., Yang, W., Bi, Y., & Shen, G. Chain breakage in liquid sulfur at high pressures and high temperatures. *Phys. Rev. B* **89**, 174201 (2014).

25. Plašienka, D., Cifra, P. & Martoňák R. Structural transformation between long and short-chain form of liquid sulfur from ab initio molecular dynamics. *J. Chem. Phys.* **142**, 154502-512 (2015).

26. Mezouar, M. et al. Development of a new state-of-the-art beamline optimized for monochromatic single-crystal and powder X-ray diffraction under extreme conditions at the ESRF. *J. of Synch. Rad.* **12,** 659-664(2005).

27. Crichton W.A., Mezouar M., Grande T., Stølen S., Grzechnik A. Breakdown of intermediate-range order in liquid GeSe2 at high pressure. *Nature* **414**, 622-625(2001)

28. Katayama, Y. et al. Density measurements of liquid under high pressure and high temperature. *J. Synchrotron Rad.* **5**, 1023-1025 (1998).

29. Bathia A.B. & March N.H. Relation between principal peak height, position and width of structure factor in dense monatomic liquids. *Phys. Chem. Liq.* **13**,313 (1984).

30. Ponyatovsky, E. G. & Barkalov, O. I. Pressure-induced amorphous phases. *Mater. Sci. Rep.* **8**, 147-191(1992).